\documentclass[12pt]{iopart} 
 
\usepackage{graphicx} 
 
\begin{document} 
 
\title{Periodically-driven cold atoms: the role of the phase} 
\author{K. Kudo$^1$ and T.S. Monteiro$^2$} 

\address{$^1$ Division of Advanced Sciences, Ochadai Academic Production, 
Ochanomizu University, 2-1-1 Ohtsuka, Bunkyo-ku, Tokyo 112-8610, Japan}
\address{$^2$ Department of Physics and Astronomy, 
University College London, Gower Street, London WC1E~6BT, UK}
\eads{$^1$\mailto{kudo.kazue@ocha.ac.jp}, $^2$\mailto{t.monteiro@ucl.ac.uk}}
 
\begin{abstract} 
Numerous theoretical and experimental studies have investigated the 
dynamics of cold atoms subjected to time periodic fields.
Novel effects dependent on the amplitude and frequency of the 
driving field, such as Coherent Destruction of Tunneling have been identified
and observed. 
However, in the last year or so, three distinct types of experiments have demonstrated
for the first time, interesting behaviour associated with the
the {\em driving phase}: i.e. for systems experiencing a driving field 
of general form 
$V(x)\sin (\omega t + \phi)$, different  types of large scale oscillations and directed motion
were observed.
We investigate and explain the phenomenon of Super-Bloch Oscillations (SBOs)
in relation to the other experiments and address the role of initial phase in general.
We analyse and compare the role of $\phi$ in systems with homogeneous forces 
($V'(x)= const$), such as cold atoms in
shaken or amplitude-modulated optical lattices, as well as
non-homogeneous  forces ($V'(x)\neq const$), such as the sloshing
of atoms in driven traps, and clarify the physical origin of the
different $\phi$-dependent effects.
\end{abstract} 
\maketitle 

\section{Introduction} 

Experiments on cold atoms in optical lattices subject to time-periodic
 perturbations \cite{Arimondo,Arimondo1,Kierig,Eck09,Arimondo2}
have provided exceptionally clean
demonstrations of  coherent phenomena predicted in
 theoretical studies
\cite{DL,CDT,CDT1,Grifoni,Holt,Holt2000,Holt2008,Garreau,Garreau1}.
Of especial note are the phenomena termed Coherent Destruction 
of Tunnelling (CDT)~\cite{CDT,CDT1,Grifoni} or Dynamic Localization 
(DL)~\cite{DL}.

In particular, if an oscillating potential $V(x,t)= Fx \cos \omega t$
is applied, it was found that the tunnelling amplitudes 
$J$ of the driven atoms take an effective, renormalized, value: 
\begin{equation} 
J_{\rm eff}\propto J \mathcal{J}_0\left(\frac{Fd}{\hbar\omega}\right), 
\label{renorm} 
\end{equation} 
where $d$ is the lattice constant and
$\mathcal{J}_0$ denotes a zero-th order ordinary Bessel function.
In the CDT ($\omega \gg J$) regime, there is a complete cessation of
tunnelling at the zeros of the Bessel function; in the DL regime, the atoms 
make a periodic excursion but return to their original position
if $\mathcal{J}_0 \sim 0$ \cite{NOTE}.\\ 

The above showed that even the one-particle dynamics can exhibit
non-trivial effects arising from matter-wave coherence.
 The potentials were realized with ultracold 
atoms in shaken optical lattices; Eq.(\ref{renorm}) was demonstrated 
and investigated experimentally in a series of ground-breaking
experiments \cite{Arimondo,Kierig}. The effects of interactions have
also been investigated ~\cite{Arimondo2,Malomed,Malomed1,Holthaus,Creffield}
such as AC-control of the Mott-Insulator to Superfluid phase transition,
including even triangular optical lattices \cite{Sengstock}.

The experiments above identified effects which depend on the frequency
and amplitude of the driving field, not its initial phase. 
However, in the last year or so, three distinct experiments on periodically
driven cold atoms have identified effects due to the phase; in other words,
for driving of form $V(x,t) \propto \sin (\omega t + \phi)$, the dynamics
was found to depend strongly on $\phi$.
For example, for atoms in shaken lattices, \cite{Ferrari,Nagerl}
 were able to realize both directed motion as well as large oscillations,
occurring over hundreds of sites, which were termed
``Super-Bloch Oscillations'' (SBOs) \cite{Nagerl}. These SBOs were
analyzed in \cite{Kolovsky}, including also the effects of
weak interactions
(mean-field regime). The general assumption in theoretical analyses has been that
 the external field-dependence of the wavepacket mean group velocity in these cases
  is entirely contained in the Bessel function argument, with no dependence on
$\phi$. However, the experiment indicates that there is a strong dependence on
this phase. 

In a second example, \cite{Alberti}, the tunneling amplitude of the
lattice was modulated. The result was a global motion (non-zero velocity
of the atomic wavepacket), strongly dependent on $\phi$. 
The third example did not involve optical lattices or
tunneling but looked at a BEC ``sloshing'' in a time-averaged orbiting 
potential (TOP) trap. Experiments showed
\cite{Wal} using a 2-dimensional set up, that the amplitude of
oscillations of atoms in traps has a strong dependence on phase.
Related theoretical work on the atoms oscillating in traps was carried out 
by \cite{Rid,Rid1}.

The atoms in traps are also driven with potentials of the form
 (in 1-dimension) $V(x) \sin (\omega t +\phi)$. Yet they differ
in essential ways. Because the
potential is not linear, the force is not uniform; the time-averaged
motion becomes strongly coupled to the phase; the trap analysis 
always requires application of an adiabatic separation between fast $\omega$ and slow
mean motion. In addition, they also differ from the optical lattice
systems because the latter are band-Hamiltonians, with an effective kinetic
energy $-J \cos p$ (in a classical-image Hamiltonian). 
Such differences mean that effects due to $\phi$  are more subtle in 
the lattices; thus for the latter,
 phase effects have been  neglected for several years,
 while they  quickly became apparent in the trap systems. 

However, at present, the physical role of the initial phase in all these experiments does not seem
too clear. Recent analyses consider the effect of the phase 
in terms of a shift in the origin of time for shaken lattices \cite{Eck09} 
or that the effects arise due to the abrupt amplitude
jump which occurs when the field is initially switched-on \cite{Wal} at $t\simeq 0$.

We have recently found \cite{SBO}, that in order to explain the experiments \cite{Ferrari,Nagerl},
 a phase correction
$\Phi(F)=F_\omega \cos \phi-n(\phi+\frac{\pi}{2})$ must 
be considered 
in an effective dispersion relation. This explains a  field dependent
shift in the phase of SBOs. It accounts also
for a separate regime of directed motion. Here we analyse 
 these SBOs in relation to other comparable
experiments with cold atoms which have also observed other types of global
motion and large scale oscillations
such as trap oscillations or amplitude-modulated
dynamics. More broadly, the aim  of this work is to clarify the role
 of the initial phase in such systems
and to consider the validity of proposed models and assumptions in these related
experiments.

It is worth stressing that for shaken lattices and SBOs, 
the strongest effects occur when the system evolves 
from $p=0$ with a smooth pure $\sin \omega t$ drive, where there is no
 amplitude jump or even a phase-jump at $t=0$.
Yet below we consider only oscillations where the phase is
well-defined in each oscillation. Thus $\phi$ has to be set on a timescale fast compared
with $T=2\pi/\omega$.  Variants with a slower ramping-up procedure 
depend on the particular protocol and we do not consider these.

Below we consider systems with homogeneous forces ($V'(x)= const$) 
separately from those for which the forces are position dependent.

\section{Inhomogeneous systems: Atoms in Traps} 

We consider first the classical dynamics for a Hamiltonian of form:
\begin{equation}
H(x,t)=H_0(p) + V(x) F(t).
\label{HAM}
\end{equation}

Here $F(t)=F(t+T)$ is a time periodic driving term and
 $H_0(p)=\frac{p^2}{2m}$ for traps or $H_0(p)=-J \cos p$ (for a
one-body image Hamiltonian for a particle in a band).
One may also consider an additional time independent 
potential, i.e. $H_0(p,x)= H_0(p) +V_0(x)$ but for the essential
physics, this is not necessary for now.

Thus there is a classical force:
\begin{equation}
{\dot p} = -V'(x) F(t).
\label{Force}
\end{equation}
If one can neglect the time evolution of $x(t)$ in the interval $[0:t]$,
we have a momentum shift on the initial momentum $p_0$:
 \begin{equation}
 p(t) =p_0 - V'(x_0) \int_0^tF(t)dt,
\label{Shift}
\end{equation}
where $x_0=x(t=0)$

 One instance is the situation,  well-studied
experimentally  \cite{QKR,QKR1,QKR2,QKR3,QKR4,QKR5}, 
of $\delta$-kicked cold atoms
 where an optical lattice potential is abruptly
switched-on for a very short time and where $F(t)=\Sigma_n \delta(t-nT)$.
For a single kick, and $T=1$,
 the shift is $p=p_0 - V'(x)$. There is an instantaneous impulse
applied to the atoms. It is well-known that, provided the switch on
is sufficiently fast and the pulse is of short duration, the dynamics
is insensitive to the pulse shape. However, the shift is position dependent;
for repeated kicks, this can have drastic consequences including chaos, for
which kicked atoms are a leading paradigm.

Another instance of a position-dependent driving,
corresponds to atoms in traps \cite{Rid,Rid1},
where $V(x)$ represents, for example, a harmonic trap of 
frequency $\Omega$ and   $F(t)=  \sin (\omega t+ \phi)$ represents a
sinusoidal drive. In the case $\omega \gg \Omega$ one can
consider a period-averaged mean motion for the slow oscillation in
the well, with coordinate $X(t)$, as well as a ``micromotion'' characterized
by the faster $\omega$ timescale. Adiabatic separability means that 
one may consider these motions separately.
 In \cite{Rid,Rid1} an effective shift of the
mean-momentum $P_0=m{\dot X(0)}$ was identified.  In the present notation,
one would obtain an effective  phase-dependent shift of the period-averaged momentum:
 \begin{equation}
 P_0' =p_0 +  \frac{V'(x_0)}{\omega} \cos \phi .
\label{Shift2}
\end{equation}

A mean-momentum shift 
 \begin{equation}
\Delta P(x_0)= \frac{V'(x_0)}{\omega} \cos \phi
\end{equation}
is zero for $\phi=\pi/2$ or $\phi=3\pi/2$. But it is
maximal for $ \phi= 0, \pi$ the case where the
$F(t)= \pm\sin \omega t$; for this case, the driving field starts
smoothly from zero. This clearly indicates
that this case represents essentially different physics from the
$\delta$-kicked particles. The shift is unrelated to the abruptness
of the change in the Hamiltonian amplitude at $t=0$, but rather to a ``slippage''
discussed in \cite{Rid,Rid1} between the slow mean motion and the fast
``micromotion''. It makes little sense to define period-averaged
motion on a timescale shorter than of order $T=2\pi/\omega$:
the system requires a finite time to ``notice'' that the mean
momentum has shifted; for the $\delta$-kicked particle, in
contrast, there is a real impulse, applied instantaneously on a timescale $\ll T$.
A formal way of stating this is that a $\delta(t)$ function implies a
finite impulse at $t \simeq 0$, while the Heaviside step function switch-on of a $\sin(\omega t+ \phi)$
drive gives zero impulse at $t\simeq 0$ and is thus not significant to the
dynamics.

Nonetheless, the shift $\Delta P(x_0)$ has real physical implications
as it can reduce or increase the kinetic energy of the mean motion,
depending on its sign.
It leads to an increase or decrease of the amplitude of the harmonic
oscillations of the particle in the trap (averaged over one period).
Ridinger and Davidson  \cite{Rid,Rid1}, for instance considered
these and other examples of traps (Eqs.~(15) and (16) therein).\\

These effects were demonstrated experimentally in atoms
sloshing in a TOP trap \cite{Wal}. 
The experiments however,
used a 2-dimensional set-up. Even for $\phi=0$, 
a rotating field $(B \cos \omega t,B \sin \omega t)$ 
is applied in the $(x,y)$ directions.
There is never a situation where a pure sine drive is applied and there is
always an initial ``jump'' in field amplitude in at least
one coordinate. Thus a convenient view (Ref.~\cite{Wal})
is to reset the large scale oscillation (macromotion) amplitude coordinates
at the end of the (near-instantaneous) switch-on of the field.
The macromotion is thus attributed to the sudden switch-on.
However, one should stress that
the underlying physics can  emerges only on the timescale of $T$, rather than
being due to the abrupt initial switch-on procedure.

\section{Homogeneous systems: shaken optical lattices}

The shaken optical lattices represent a different class of Hamiltonians.
The corresponding classical Hamiltonian is:
\begin{equation}
H(x,t)=-J \cos p - F x \sin (\omega t +\phi).
\label{SHAK}
\end{equation}
A key difference is that $V'(x)=F$ is independent of
position. While the particle does display Bloch
Oscillations (SBOs) with the period $T_B$, 
there is no separation of timescales since
$T_B \simeq T$. There is no restriction to high frequency driving.
Further, there is a decoupling between the instantaneous
phase of the Bloch Oscillations and
the phase, since the initial position $x(t=0)$ is immaterial: the force
is uniform. The initial wavepacket is even delocalized over several lattice sites.
From the classical equations of motion, $p=p_0+ F\int_0^t \sin (\omega t' +\phi) dt'$
and a effective momentum shift arises from the integrand at $t=0$:
\begin{equation}
 p(t)=p_0 +  \frac{F}{\omega} \cos \phi - \frac{F}{\omega}\cos ({\omega}t+ \phi).
\end{equation}
The effective shift is, however, position independent:
 $p_0'=p_0 + \frac{F}{\omega}\cos \phi$. 
It is possible to then consider the effect of the phase as simply
a corresponding displacement in the origin of time as suggested in 
\cite{Eck09}, but only
in the momentum shifted frame of $p'_0$.

Once again, for the completely smooth switching on of the driving Hamiltonian
(pure $\sin \omega t$ drive) the shift is a maximum. 
And here it is not useful to think of an initial ``jump'' either in
amplitude or phase. Taking the
experimental value $p_0=0$ and $\phi=0$,

\begin{equation}
\langle p(t)\rangle =  \frac{F}{\omega}\langle 1  -\cos {\omega}t\rangle .
\end{equation}

We see that the momentum now oscillates about a shifted non-zero value,
but the process is not related to the abruptness of the switch-on,
 which is perfectly smooth. The shift in the average momentum becomes apparent
 only on the timescale of $T$. This is very different from the $\delta$-kicked
 particle, where a real instantaneous impulse is imparted. It is also different from
 the trap, where through $V'(x_0)$, the position at
the instant the phase is set is important.\\

One may solve Hamilton's classical equations of motion to obtain a period averaged
group velocity:
\begin{equation}
v_g \simeq -J \mathcal{J}_0\!\!\left(F_\omega\right)\sin \left(p+F_\omega\cos \phi \right),
\end{equation}
where $F_\omega=F/\omega$.

In the shaken lattice, there is no dependence on initial position $x_0$. Thus we conclude that
 the directed motion is not related to the sudden
switch-on of the field. It is best understood as a process more akin
to ratchet physics and the de-symmetrization of the drive. The
time-asymmetric $\sin \omega t$ drive generates a ratchet current
(since the average force is zero, there is no net bias: thus this
fulfils one of the criteria for ratchet motion, though not the stricter one of
 ``rectification of fluctuations'', so should perhaps be termed simply
``directed motion''). The symmetric drive $\cos \omega t$ produces 
no ratchet motion, while intermediate values of $\phi$ produce intermediate
 behaviour. The final current is similar to that obtained in the linear-ramping study
of Ref.\cite{Creff2}.

The above discussion explains only directed motion and its $\phi$ dependence.
For Super-Bloch Oscillations (SBOs), one has an additional static
field. One must consider in addition the quantum resonance
between $\omega$ and the gap between energy levels in adjacent wells.
We thus consider a quantum treatment, within a Floquet-theory framework
(full details are in \cite{SBO}).

 We consider the total Hamiltonian:
$H(t) = H_0 +H_F(t)$, where
\begin{equation}
H_0=-\frac{J}{2}\sum_{j,\sigma}(c_{j,\sigma}^\dagger c_{j+1,\sigma}+ \mathrm{H.c.}).
\end{equation}
 $H_0$ corresponds to the non-interacting limit of
a variety of Hamiltonians with nearest-neighbour hopping (Hubbard, Bose-Hubbard,
magnons in Heisenberg spin chains, etc.)
It represents any
spatially periodic potential characterized by energy eigenfunctions $\phi_{mk}$,
with band index $m$ and wavenumber $k$,
thus $H_0 \phi_{mk} = E_m(k)  \phi_{mk}$.
We restrict our one-particle problem  to the
lowest band $m=1$; taking $E(k) \equiv E_{m=1}(k)$, the energy dispersion
$ E(k) = -J \cos kd$.

We take
$H_F(t) = -f(t) x$ 
where $f(t)=F_0+ F \sin (\omega t+\phi)$; in general it comprises both a static field
 and a sinusoidally oscillating field with an arbitrary phase $\phi$.
The result of
 the driving is a time-dependent wavenumber \cite{Eck09}:
\begin{equation}
q_k(t)=k +\frac{1}{\hbar} \int_0^t d\tau f(\tau)=
 k+ \frac{F}{\omega} \cos \phi - \frac{F}{\omega}\cos ({\omega}t+ \phi)
\label{waveq}
\end{equation}

The stationary states of the
 system are its Floquet states, the analogues of Bloch waves in a temporally
 periodic system. They
are given by
$\psi_k(x,t)=u_k(x,t) \exp\left[-\frac{i}{\hbar} \epsilon(k) t \right]$,
where $u_k(x,t)=u_k(x,t+T)$, and the period $T=2\pi/\omega$.
The non-periodic phase-term is characterized by
the so-called ``quasienergy''  $\epsilon(k)$.

The presence of the static linear field in general destroys the 
band dynamics; however, here we consider the so-called
resonant driving case, for
 which $F_0 d= n \hbar\omega$, 
(where the driving compensates for the energy offset
between neighbouring wells in the
 lattice, restoring the band structure). In this case, other
studies, while neglecting $\phi$, have found that
the renormalization of tunnelling is $n-$dependent:
\begin{equation}  
J_{\rm eff}\propto J \mathcal{J}_n\left(\frac{Fd}{\hbar\omega}\right),
\label{renorm_n} 
\end{equation} 
where $d$ is the lattice constant and
$\mathcal{J}_n$ denotes an ordinary Bessel function
and $n=0,1,2..$.
Below we take $\hbar=1$ and $F_\omega=Fd/\omega$.

For the purposes of calculating to group velocities, 
we define  an effective quasienergy dispersion:
\begin{equation}
\epsilon(k) = -\frac{J}{T} \int^T_0 \cos (q_k(t)d) dt
\label{quasi}
\end{equation}
by means of a period-average.

Below we also consider the case of slight detuning for which
$F_0 d= (n+ \delta) \omega$, with $\delta \ll 1$, associated 
with SBOs; however,
 for the slight-detuning case $\delta \neq 0$, we assume that
the time-dependence due to the $\delta \omega t/d$ remains negligible
over one period $T$. Thus we take it out of the period-average
entirely and, as shown in \cite{SBO},  Eq.~(\ref{quasi}) becomes:
\begin{equation}
\epsilon(k) \simeq 
-J\mathcal{J}_n(F_\omega) \cos \left[ kd + \delta \omega t
+F_\omega \cos \phi  - n(\phi + \frac{\pi}{2})\right].
\label{dispersion}
\end{equation}
 The above represents an {\em effective} dispersion
relation, but which {\em oscillates slowly in time}  with a period
$T_{\rm SBO}=2\pi/(\delta\omega) \gg T_B$ where $T_B \propto 1/F_0$
is the Bloch period.
They correspond to the ``Super-Bloch Oscillations'' investigated by
Refs.~\cite{Garreau,Garreau1,Kolovsky,Ferrari,Nagerl}. Even at resonance
$ \delta \omega=0$,
Eq.~(\ref{dispersion}) differs from previous expressions by the phase-shifts
$\Phi(F)= F_\omega \cos \phi - n(\phi + \frac{\pi}{2})$.
Evaluating Eq.~(\ref{dispersion}) for the $n=1$ case, we obtain a period-averaged
group velocity:
\begin{equation}
v_g= \frac{\partial \epsilon}{\partial k}\simeq -Jd 
\cos \left(kd+F_\omega\cos \phi -\phi + \delta \omega t\right)  \mathcal{J}_1\!\!\left(F_\omega\right).
\label{VG}
\end{equation}


\begin{figure}[htb] 
\begin{center}
 \includegraphics[height=1.5in]{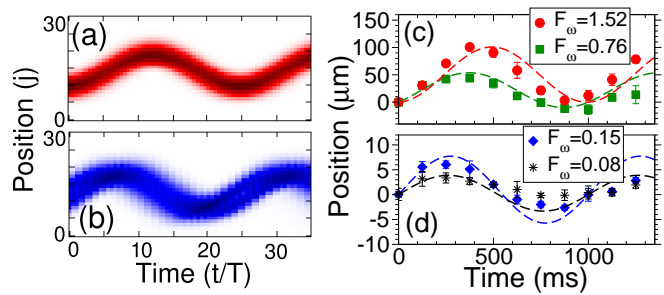} 
\includegraphics[height=1.5in]{figure1c_1d.eps} 
\end{center}
\caption{ {\bf (a)} and {\bf (b)}
One-particle solutions of the Hubbard
Hamiltonian with the driving term 
for $L=30$ lattice sites, showing the field-dependence
of the phase of Super-Bloch Oscillations~\cite{SBO}.
Upper/lower panel  correspond to  $F_\omega=1.52$ and $0.15$;
both have $\phi=\pi$.
For illustration purposes, the theoretical amplitudes were equalized
by choosing $J$-values
which equalize $J_{\rm eff} \simeq J \mathcal{J}_1(F_\omega)$.
{\bf (c)} and { \bf (b)} Show that good agreement is
obtained between experiment
and the Hubbard numerics, as well as the analytical formula~\cite{SBO}:
integrating Eq.~(\ref{VG}) (dashed lines)
reproduces well both the amplitude and phase of SBO 
 experimental data (symbols) of  Ref.~\cite{Nagerl}, using
$\delta\omega=2\pi/1000$ ($ \equiv -1$ Hz detuning),
 $\phi=\pi$ and $Jd/(\hbar\delta\omega)=90$ $\mu$m.}
\label{Fig1} 
\end{figure}
                   
The above expression accounts for a range of experimental features.
However, the most interesting one is that the SBOs begin with a
field-dependent phase.
Figure~\ref{Fig1} demonstrates this for experimental field values
$F_\omega=0.08$, $0.15$, $0.76$ and $1.52$ (compared with one-particle
Hubbard numerics). The graphs  show that
Eq.~(\ref{VG}) reproduces quite well 
the experimental values of Ref.~\cite{Nagerl}.
They show clearly the displacement of the first maximum,
seen in the experiment as well
as the order of magnitude variation in amplitude.
Such a field-dependent shift is also
apparent in the results of Ref.~\cite{Ferrari}. \\

Finally, we consider the case of experiments with amplitude modulated lattices,
 which do not show renormalisation of tunneling, but
 can also produce directed motion \cite{Alberti}.
The experiments also used a static linear field $F_0x$.
An experimental group velocity $ v_g \propto J \cos \left(kd -\phi\right)$,
for the $n=1$ resonance case $F_0d=\hbar \omega$,  
was measured.
Curiously, the effective momentum shift $F_\omega\cos \phi$
seen in other driving systems is no longer present.

We find that the observed behavior in \cite{Alberti} is consistent with
an effective classical Hamiltonian for the homogeneous force system:
\begin{equation}
H(x,t)=-J[1+ \alpha\sin (\omega t +\phi)] \cos p + F_0x.
\label{Alb}
\end{equation} 
Solving for Hamilton's classical equations of motion, we see
that $p(t)=p_0- F_0t$ no longer has a phase dependence. 
For simplicity, we take $\hbar=d=1$ and thus $F_0=\omega$.
Then we easily see that the period averaged velocity 
$\langle\Delta x \rangle/T= \frac{1}{T}\int_0^T {\dot x(t)} dt= 
-\frac12 J\alpha \cos (p_0-\phi)$.
In other words, even with  zero momentum shift, in this case
a non-zero phase correction can arise.

\section{Conclusion}

In this study, we have attempted to clarify the role of driving
phase in the dynamics of homogeneous and inhomogeneous-force systems.
To this end we have compared recent experimental results
on homogeneous systems (shaken lattices) and inhomogeneous
systems (atoms oscillating in traps). 

It is tempting to draw analogies between the ``slow'' SBOs
 with the macromotion or 
``sloshing'' seen in traps \cite{Wal}. However, in the latter,
the effect of the phase is manifested directly in the amplitude
of the oscillations in the trap. In the shaken lattices, the manifestation
of the phase is more subtle. The amplitude of SBOs is
independent of $\phi$, only their phase is affected.
Unlike the trap systems, their time-dependence
does not arise from the period-average integral.
Nonetheless, it seems clear that in order to account for the
experiments it is necessary to include the phase-correction
$\Phi(F)$ in an effective dispersion, used to obtain group velocities.

What all these systems do have in common is that the main effect of the
initial phase is to modify the global centre of mass dynamics of the wavepacket.
This is in contrast to previous experiments on, for example, DL, which
showed diffusive spreading of the wavepacket rather than large-scale
motion, on scales of many sites.

\ack

We acknowledge helpful comments or discussions with A.~Eckardt,
M.~Holthaus and C.~Weiss.
We are extremely grateful to
E.~Haller and C.~N\"agerl for the experimental
data in Fig.~\ref{Fig1}.
This work is partly supported by KAKENHI(21740289).


\section*{References}

 \begin{thebibliography}{99} 
\bibitem{Arimondo} Lignier H, Sias C, Ciampini D, Singh Y, Zenesini A, 
Morsch O and Arimondo E 2007 \PRL {\bf 99} 220403 
\bibitem{Arimondo1} Sias C, Lignier H, Singh Y, Zenesini A, Ciampini D, Morsch O
and Arimondo E 2008 \PRL {\bf 100} 040404
\bibitem{Kierig} Kierig E, Schnorrbrber U, Schietinger A, Tomkovic J
and Oberthaler M  2008 \PRL {\bf 100} 190405
\bibitem{Eck09} Eckardt A, Holthaus M, Lignier H, Zenesini A, Ciampini D,
Morsch O and Arimondo E 2009 \PR A. {\bf 79} 013611
\bibitem{Arimondo2}  Zenesini A, Ligier H, Ciampini D, Morsch O and 
Arimondo E 2009 \PRL {\bf 102} 100403
\bibitem{DL} Dunlap D H and Kenkre V M 1986 \PR B {\bf 34} 3625
\bibitem{CDT} Grossmann F, Jung P, Dittrich T and Hanggi P 1991 \ZP B 
{\bf 84} 315
\bibitem{CDT1} Grossmann F, Dittrich T, Jung P and Hanggi P 1991 \PRL 
{\bf 67} 516
\bibitem{Grifoni} Grifoni M and Hanggi P 1998 {\it Phys. Rep.} {\bf 303} 229
\bibitem{Holt} Holthaus M 1992 \PRL {\bf 69} 351
\bibitem{Holt2000} Holthaus M 1999 \textit{Coherent control in Atoms,
Molecules and Semiconductors}
 eds Potz W and Schroeder W A (Dordrecht: Kluwer) pp. 171-182
\bibitem{Holt2008} Eckardt A and Holthaus M  2008 
{\it J. of Phys.: Conf. Series} {\bf 99} 012007
\bibitem{Garreau} Thommen Q, Garreau J-C, Zehnle V 2002 \PR A
{\bf 65} 053406
\bibitem{Garreau1} Thommen Q, Garreau J-C, Zehnle V 2004 \JOB {\bf 6} 301
\bibitem{NOTE} We use the term DL-regime loosely to mean
there is no restriction to high $\omega$, rather than the exact fields
for which there is a zero of a Bessel function.
\bibitem{Malomed} Mayteevarunyoo T and Malomed B A 2009 \PR A  
 {\bf 80} 013827 
\bibitem{Malomed1} Mayteevarunyoo T, Malomed B A and Roeksabutr A 2010 
\JOSA {\bf 27} 1957
\bibitem{Holthaus} Eckardt A, Weiss C and Holthaus M 2005 \PRL 
 {\bf 95} 260404
\bibitem{Creffield} Creffield C E and Monteiro T S 2006 \PRL
{\bf 96} 210403
\bibitem{Sengstock} Eckardt A, Hauke P, Soltan-Panahi P, Becker C,
Sengstock K and Lewnstein M 2010 {\it Europhys. Lett.} {\bf 89} 10010
\bibitem{Ferrari} Alberti A, Ivanov V V, Tino G M and Ferrari G
2009 {\it Nature Physics} {\bf 5} 547
\bibitem{Nagerl} Haller E, Hart R, Mark M J, Danzl J G, Reichs\"ollner L
and N\"agerl H-C, 2010 \PRL {\bf 104} 200403
\bibitem{Kolovsky} Kolovsky A R and Korsch H J 2010 \PR A
{\bf 82} 011601R
\bibitem{Alberti} Alberti A, Ferrari G, Ivanov V V, Chiofalo M L and
Tino G M 2010 \NJP {\bf 12} 065037
\bibitem{Wal} Cleary P W, Hijmans T W and Walraven J T M 2010 \PR A 
{\bf 82} 063635
\bibitem{Rid} Ridinger A and Davidson N 2007 \PR A {\bf 76} 013421
\bibitem{Rid1} Ridinger A and Weiss C 2009 \PR A {\bf 79} 013414 
 
\bibitem{SBO} Kudo K and Monteiro T S 2011 {\it Preprint} arXiv:1102.0599

\bibitem{QKR} Moore F L, Robinson J C, Bharucha C F, Sundaram B and
Raizen M G 1995 \PRL {\bf 75} 4598;
\bibitem{QKR1} Ammann H, Gray R, Shvarchuck I and Christensen N 1998 
\PRL {\bf 80} 4111
\bibitem{QKR2} Szriftgiser P, Ringot J, Delande D and Garreau J C 2002 \PRL
{\bf 89} 224101;
\bibitem{QKR3} Duffy G J, Parkins S, Muller T, Sadgrove M, 
Leonhardt R and Wilson A C \PR E {\bf 70} 056206
\bibitem{QKR4} Jones P H, Stocklin M M A, Hur G and Monteiro T S 2004 
\PRL {\bf 93} 223002
\bibitem{QKR5} Hur G, Creffield C E, Jones P H and Monteiro T S 2005 
\PR A {\bf 72} 013403

\bibitem{Creff2} Creffield C E and Sols F 2008 \PRL {\bf 100} 250402
\bibitem{Kudo} Kudo K, Boness T and Monteiro T S 2009 \PR A {\bf 80} 063409
\end{thebibliography}
\end{document}